\documentclass[12pt]{iopart}
\pdfoutput=1
\usepackage[T1]{fontenc}
\usepackage[latin9]{inputenc}
\usepackage{amssymb}
\usepackage{graphicx}
\usepackage{color}

\newcommand{\qq}{\begin{eqnarray}}
\newcommand{\qqq}{\end{eqnarray}}
\newcommand{\p}{\partial}

\newcommand{\bfJ}{{\bf J}}
\newcommand{\bfq}{{\bf q}}

\begin{document}

\title[From bulk to microphase separation in scalar active matter]{From bulk to microphase separation in scalar active matter: A perturbative renormalization group analysis}

\author{Fernando Caballero} 
\address{DAMTP, Centre for Mathematical Sciences, University of Cambridge, Wilberforce Road, Cambridge CB3 0WA, UK}
\ead{fmc36@cam.ac.uk}

\author{Cesare Nardini} 
\address{Service de Physique de l'\'Etat Condens\'e, CNRS UMR 3680, CEA-Saclay, 91191 Gif-sur-Yvette, France}
\ead{cesare.nardini@gmail.com}

\author{Michael E. Cates} 
\address{DAMTP, Centre for Mathematical Sciences, University of Cambridge, Wilberforce Road, Cambridge CB3 0WA, UK}

\vspace{10pt}
\begin{indented}
\item[]\today
\end{indented}

\begin{abstract}
We consider a dynamical field theory (Active Model B+) that minimally extends the equilibrium Model B for diffusive phase separation of a scalar field,
by adding leading-order terms that break time-reversal symmetry. It was recently shown that such active terms can cause the bulk phase separation of Model B to be replaced by a steady state of microphase separation at a finite length scale. This phenomenon was understood at mean-field level as due to the activity-induced reversal of the Ostwald ripening mechanism, which provides the kinetic pathway to bulk phase separation in passive fluids. This reversal occurs only in certain ranges for the activity parameters. In this paper we go beyond such a mean-field analysis and develop a $1$-loop Renormalisation Group (RG) approach. We first show that, in the parameter range where bulk phase separation  is still present, the critical point belongs formally to the same (Wilson-Fisher) universality class as the passive Model B.  In contrast, in a parameter range associated with microphase separation, we find that an unstable non-equilibrium fixed point of the RG flow arises for $d\geq 2$, colliding with the Wilson-Fisher point in $d\to 2^+$ and making it unstable in $d=2$. At large activity, the flow in this region is towards strong coupling. We argue that the phase transition to microphase separation in active systems, in the physically relevant dimensions $d=2$ and $3$, very probably belongs to a new non-equilibrium universality class. Because it is governed by the strong-coupling regime of the RG flow, our perturbative analysis leaves open the quantitative characterization of this new class.
\end{abstract}

%\maketitle

\section{Introduction}
Universal properties play a crucial role in theoretical physics, providing cases where the precise analysis of a minimal model gives accurate -- and non-trivial -- information about much more complex systems. In equilibrium systems, one of the most famous universality classes is that of $\phi^4$ theory describing vapor-liquid phase separation in the proximity of the critical point where the two phases become identical. The same theory describes the critical behavior of a binary fluid mixture close to its demixing transition, as well as the transition between paramagnetic and ferromagnetic states in the Ising model of ferromagnetism~\cite{goldenfeld2018lectures}. While sharing a universality class for equilibrium statistics, $\phi^4$ systems have several classes for their dynamics depending on whether the order parameter $\phi$ is conserved (true for fluids but usually not for magnets) and also whether momentum is conserved (true for bulk 3D fluids but not for quasi-2D fluid films supported by a momentum-absorbing wall). 

Since the introduction of dynamical Renormalization Group (RG) techniques~\cite{Forster1977}, universality classes in non-equilibrium systems have also been classified and studied in depth~\cite{Tauber2014}. 
A well defined and important subset of non-equilibrium systems comprise active materials, in which individual particles continually consume a fuel source in order to self-propel~\cite{ramaswamy2017active,Marchetti2013RMP}. Active systems are widespread and varied, so that a number of different universality classes are needed to describe them.
Toner and Tu~\cite{toner1998flocks,toner1995long} studied self-propelled particles, without momentum conservation, whose interactions cause them to align their directions of motion with those of their neighbors. In the `flocking' phase they found that the number of particles in a mesoscopic box undergoes giant fluctuations, with an universal exponent, and that spatial and temporal correlations decay algebraically. 
Other classes describe active nematics (with or without momentum conservation)~\cite{ramaswamy2003active,mishra2010dynamic,Marchetti2013RMP}, incompressible flocks~\cite{chen2015critical,chen2016mapping} and chemically interacting and dividing particles~\cite{gelimson2015collective}. 

Arguably the simplest active systems involve self-propelled particles without alignment interactions but with a finite relaxation time for random rotation of the swimming direction. 
(In the case where the rotational dynamics is Brownian diffusion, these are called Active Brownian Particles or ABPs.) At large length and time scales the system can then be described by a scalar density field, which is dynamically conserved. We call these `scalar active' systems. The isotropic interactions between particles can be simple pairwise forces (responsible for inter-particle collisions), and/or represented by a local dependence of propulsion speed on density. 
Somehow surprisingly, with the exception of a very recent computational work~\cite{siebert2017critical}, the universal critical properties of scalar active systems have not yet been analysed.  However, much is known phenomenologically (and from mean-field theory) about their phase equilibria well away from any critical regime. Specifically, broken time-reversal symmetry inherent in the microscopic definition of active systems, can lead to various features impossible in thermal equilibrium. 

One of these is motility-induced phase separation (MIPS) where an assembly of repulsive, but active, particles phase separates into bulk dense and dilute regions~\cite{Tailleur:08,Cates:15,Fily:12}. In some cases, the kinetics of this phase separation closely resembles that of a passive system with attractive interactions, allowing the possibility that time-reversal symmetry (TRS) might be recovered upon coarse-graining~\cite{Tailleur:08,Cates:15}, thus mapping MIPS to an effective equilibrium system at large spatial and time scales. If exactly true, this would imply that the critical transition from homogeneous to phase separated states in active systems belongs to the same universality class of passive Model B in the classification of Hohenberg and Halperin~\cite{hohenberg1977}. In particular, adding the simplest TRS-breaking term $\lambda(\nabla\phi)^2$ to the chemical potential to the Model B dynamics (creating a model called Active Model B) changes the coexisting densities but does not qualitatively alter the phase-separation dynamics ~\cite{wittkowski2014scalar,stenhammar2013continuum,solon2018generalized}.

However, in simulations of purely repulsive particles~\cite{Stenhammar14} and in experiments with synthetic self-propelled colloids~\cite{Palacci:12,Speck:13,thutupalli2017boundaries}, physics resembling microphase separation is also seen in some regions of parameter space. This can be either in the form of a dynamic population of dense clusters in a dilute sea~\cite{Palacci:12,Speck:13,thutupalli2017boundaries}, or dilute vapor bubbles in a liquid~\cite{Stenhammar14}. More specifically, the simulations of~\cite{Stenhammar14} show that purely repulsive ABPs can undergo bulk phase separation between a vapor phase and a microphase separated state, the latter composed of a dense liquid which supports in its interior dilute vapor bubbles. This closely resembles the state shown in Fig. \ref{fig:phenomenology-AMB+}(b) below.

Very recently~\cite{tjhung2018reverse}, it has been argued that microphase separation can be expected generically in many scalar active systems, without any need to invoke system-specific details~\cite{Farage:2015:PRE,alarcon2017morphology,prymidis2015self,mani2015effect,mognetti2013living} or the presence of long-range interactions~\cite{tiribocchi2015active,matas2014hydrodynamic,thutupalli2017boundaries,liebchen2015clustering,saha2014clusters}. This is due to the fact that macroscopic currents that break time-reversal symmetry can reverse Ostwald ripening, thus arresting phase separation to a finite length scale, independent of the system size. Refs.~\cite{Nardini2017,tjhung2018reverse} introduced and analysed a field-theory that 
extends equilibrium Model B to include nonlinearities up to the first nontrivial order (in the sense that TRS is broken) in a gradient expansion.
The resulting model was named Active Model B+ (AMB+) and is defined by equations (\ref{eq:AMB+})-(\ref{eq:AMB+mu}). In~\cite{tjhung2018reverse}, the phase diagram of AMB+ was studied far from any critical point by analytical mean-field arguments and numerical simulations. 

In this paper we go beyond the mean-field analysis of~\cite{tjhung2018reverse} and study the critical-point behavior of phase separation in active systems by applying one-loop dynamical RG to AMB+. 
For dimension $2\leq d <4$, we argue that the transition from homogeneous to bulk phase separation belongs formally to the universality class of equilibrium Model B, 
while the transition to microphase separation represents a new nonequilibrium universality class. The latter result is surprising from a technical point of view, because any nonlinearity that can be added to Model B in order to both break TRS and respect mass conservation should be expected to be irrelevant from dimensional analysis. Our one-loop computation however shows that a new fixed point in the RG flow arises, which is repulsive along a single direction, given by a linear combination of the active couplings $\lambda$ and $\zeta$ that requires the latter to be nonzero. Beyond a separatrix in the space of these couplings, the one-loop RG flow is towards strong coupling. This regime is closely connected with one recently reported for a nonequilibrium surface growth model (which is actually obtained from AMB+ by removing all non-gradient terms from the free energy)~\cite{caballero2018strong}. As there, while our perturbative calculation cannot definitively identify the resulting behavior, the most likely interpretation is that we have a new class whose existence is only possible because of the presence of active terms in the equations of motion. If so, the idea that TRS could be restored at large scales in active systems remains a possibility near the critical point for bulk phase separation but not in other regions of the phase diagram of equally general interest. Careful investigation is needed~\cite{Nardini2017} and indeed we show elsewhere that even near the Wilson-Fisher (model B) fixed point, where active terms are irrelevant in the RG sense, they might still lead to nontrivial steady-state entropy production~\cite{caballeroEntropy}.

The rest of this paper is organised as follows. In Section \ref{sec:model}, we introduce the model of interest (AMB+) and recall its phenomenology, summarising the results obtained at mean-field level and using numerical simulations in~\cite{tjhung2018reverse}. In Section \ref{sec:RG-intro} we outline the RG calculation and arrive at the main technical results of this paper, namely the RG flow equations of AMB+ to one loop. This section is mainly technical and can be skipped in a first reading.
We then discuss the physics that can be gleaned from these calculations in Section~\ref{sec:nu0}, by focusing on a case where we neglect a specific activity-induced nonlinearity. (The latter arises as a cross-term between the explicit activity terms in $(\lambda,\zeta)$ and the $\phi^4$ nonlinearity that is already present in passive Model B.) This approximation allows us to obtain analytical, and easily interpretable, results. In Section \ref{sec:full-RG-flow} we numerically study the full one-loop RG flow of AMB+ and thereby  confirm the main conclusions drawn from the approximate treatment in Section~\ref{sec:nu0}.
Some details of the more technical parts of the RG calculation are deferred to \ref{app:1a} and \ref{app:1f}. We summarize our work  and discuss future perspectives in Section \ref{sec:conclusions}.

%%%%%%%%%%%%%%%%%%%%%%%%%%%%%%%%%%%%%%%%%%%%%%%%%%%%%%%%%%%%%%%%
\section{Active Model B+}\label{sec:model}
\begin{figure}\center
	\includegraphics[scale=0.45]{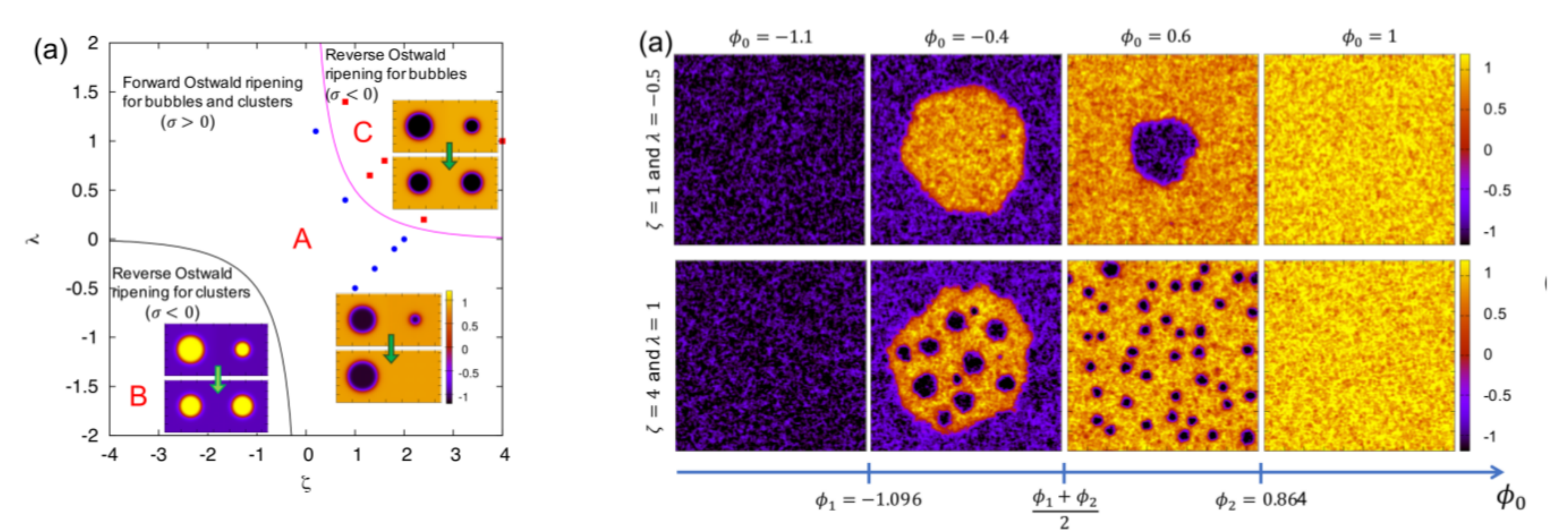}
	\caption{   Reproduced with permission from~\cite{tjhung2018reverse}: phase diagram of AMB+. In (a), analytical results within the mean-field approximation ($D=0$) predict where Ostwald Ripening is normal (region A), where it is reversed for dense clusters in a dilute environment (region B), and where it is reversed for 
	dilute bubbles in a dense environment (region C). 
	In (b) are shown numerical results obtained at finite noise ($D\neq 0$): crossing the transition line from regions A to B or from A to C corresponds to going from a bulk phase-separated state (top row, second and third panel) to a microphase-separated one (bottom row, third panel), or to a coexistence between a homogeneous state and a microphase separated one (bottom row, second panel). Parameters used are $-a=u=1/4, K=1, \nu=0=K_1$.}
	\label{fig:phenomenology-AMB+}
\end{figure}

The model we consider is AMB+, which was introduced in ~\cite{Nardini2017,tjhung2018reverse}, first on the basis of symmetry arguments, and then by explicit coarse graining of a model of self-propelled particles. It describes both bulk phase separation and microphase separation in scalar active systems. Within this model, the qualitative physics of bulk phase separation emerges quasi-passively, in the sense that the even if the underlying mechanism is purely active (as in MIPS), after coarse-graining this leads to an effective free-energy structure which has two competing minima corresponding to the dense and dilute phases. Active gradient terms play a quantitative role only, by perturbing the densities of these coexisting phases. In contrast, explicitly active terms (those that break TRS in the coarse-grained equations of motion) are essential for microphase separation within this model: the terms in the equations of motion that are compatible with an effective free energy cannot by themselves support steady-state ordering on a finite, rather than infinite, length scale. (Of course, one can construct passive models that do support microphase separation, but the passive limit of AMB+ is Model B, which does not.)

AMB+ is defined by an equation of motion for a scalar conserved density field $\phi$ (related to the physical density via a standard linear transformation~\cite{stenhammar2013continuum}):
\qq
\p_t\phi&=&-\nabla\cdot\left(\mathbf{J}+\sqrt{2DM}\mathbf{\Lambda}\right)\,,\label{eq:AMB+}\\
 {\bf J}/M &=&-\nabla \mu  + \zeta (\nabla^2\phi)\nabla\phi \,,\label{eq:AMB+J}\\
 \mu &=& \frac{\delta \mathcal{F}}{\delta\phi} +\lambda|\nabla\phi|^2 + \frac{\nu}{2} \nabla^2\phi^2\,.\label{eq:AMB+mu}
\qqq
Here $\bfJ$ is the current and $\mathbf{\Lambda}(\mathbf{r},t)$ is a Gaussian noise with zero mean and unit variance. The (nonequilibrium) chemical potential is denoted by $\mu $. This has an effectively passive part, inherited from Model B, where 
\qq
\mathcal{F} = \int \left(f(\phi) + \frac{K}{2}(\nabla\phi)^2\right)d\mathbf{r},\qquad f = \frac{a}{2} \phi^2+\frac{u}{4} \phi^4 \label{eq:AMB+F}
\qqq
where $f$ is the local thermodynamic potential, with $K>0$. The mobility $M$ is normally set to be $\phi$-independent in equilibrium ($M=1$) and we do so here too, although a $\phi$-dependent mobility $M$ arises when deriving AMB+ from explicit coarse graining of a particles model. The same applies to passive Model B, to which AMB+ reduces when $\lambda=\zeta=\nu=0$. Whereas passive Model B is symmetric under $\phi\to -\phi$, this symmetry is at first sight broken by the active terms. But in fact symmetry is not lost, merely altered: AMB+ remains symmetric under the generalized transformation $(\lambda,\zeta,\nu,\phi)\to -(\lambda,\zeta,\nu,\phi)$.

Two lines of reasoning originally motivated the introduction of AMB+. First, and similarly to what was done in passive Model B~\cite{hohenberg1977}, AMB+ describes phenomenologically a system undergoing active phase separation at leading nontrivial order in a gradient expansion, allowing all terms in $\dot \phi$ to order $\mathcal{O}(\nabla^4\phi^2)$; this is the order at which the first terms to explicitly break TRS arise. Second, it was shown in~\cite{tjhung2018reverse} that the coarse-graining of a model of self-propelled particles indeed leads to AMB+, albeit with non-constant mobility $M$ (resulting also in multiplicative noise), and with a more complex local thermodynamic potential $f(\phi)$. 

In~\cite{tjhung2018reverse}, the parameter $\nu$ in (\ref{eq:AMB+mu}) was set equal to zero, but the constant $K$ in (\ref{eq:AMB+F}) was generalized to take the form $K + 2K_1\phi$. This formulation is equivalent to ours because all possible non-linear terms at order $\mathcal{O}(\nabla^4\phi^2)$ can be written as a linear combination of either $(\lambda,\zeta,\nu)$ or of $(\lambda,\zeta,K_1)$ nonlinearities. (Note that in \cite{Nardini2017} terms of even higher order were additionally considered; these will not concern us here.) Observe that, in our representation, whenever $2\lambda = -\nu$, the chemical potential $\mu$ can be restored to an equilibrium form by setting $K \to K(\phi) = K+2\lambda\phi$ within ${\cal F}$. Accordingly, there are only two independent sources of explicit activity within the model. Note that for much of the analytic mean-field theory in~\cite{tjhung2018reverse}, the (effectively passive) nonlinearity $K_1$ was furthermore set to zero. It may seem reasonable to ignore any nonlinearity that is not explicitly active, and indeed it was shown in ~\cite{tjhung2018reverse} that doing so does not qualitatively alter the mean-field behavior. Here, however, we do not set $\nu$ to zero because this coupling can acquire a finite value under the RG flow even if zero initially. This has important consequences for the critical behavior, as will become clear below.

The phenomenology of AMB+ shows unexpected regimes of microphase separation. In passive systems undergoing diffusive phase separation without momentum conservation, coarsening of domains is driven to completion by the Ostwald process~\cite{Bray,CatesJFM:2018}. As normally understood, this process always causes the shrinkage of small domains and the growth of large ones. Instead, in the active case, the Ostwald process can go into reverse leading to stable, finite-size clusters or bubbles.  The analysis of the Ostwald process for AMB+ performed in~\cite{tjhung2018reverse} is a mean-field one, and leads to the phase diagram in Fig.~\ref{fig:phenomenology-AMB+}(a). In region A ($\lambda$ and $\zeta$ small in magnitude or of opposite signs), Ostwald ripening is normal both for dilute bubbles in a dense environment and for dense clusters in a vapor environment. The system however undergoes a phase transition with increasingly positive $\lambda, \zeta$ into region C: here,
dilute bubbles in a dense environment undergo reversed Ostwald ripening, where the smaller ones grow at the expense of larger ones, which shrink. (In contrast, dense clusters in a dilute environment undergo normal Ostwald ripening in this region.) Finally, region B is obtained from C exploiting the symmetry of the model $(\lambda,\zeta,\nu,\phi)\to -(\lambda,\zeta,\nu,\phi)$. Here the reverse Ostwald process is seen for dense clusters in a dilute vapor.

The effect of noise in AMB+ was so far investigated via numerical simulations only. It was shown that, crossing the transition line between A and C regions, bulk phase coexistence is transformed into microphase separation, or into phase coexistence between a uniform state and a microphase-separated one; see Fig. \ref{fig:phenomenology-AMB+}(b) where the numerically obtained steady state is shown for different values of the average density $\phi_0$. In region C the microphase-separated state is formed of dilute bubbles while in region B it is formed of dense clusters.

While TRS is always broken at the level of the equations of motion (unless both $2\lambda=-\nu$ and $\zeta=0$) no clear large-scale signature of this irreversibility has yet been identified for systems within region A of the phase diagram in Fig. \ref{fig:phenomenology-AMB+}(a). The bulk coarsening dynamics seen here numerically instead appears qualitatively very similar to that of passive Model B. In contrast, in the steady states featuring microphase separation (regions B and C), time-reversal symmetry is manifestly broken at or beyond the length scale of the emergent structures. This is particularly clear when noise is low: here bubbles (clusters) are created by nucleation but disappear either because of coalescence with other bubbles (clusters) or because they are ejected out into a coexisting bulk vapor (liquid) phase -- see Supplementary movie 4 in~\cite{tjhung2018reverse}.

To summarize the above, AMB+ shows an interesting variety of phenomena including, but not limited to, activity-driven microphase separation. While some elements of this behavior, such as reverse Ostwald ripening, can be understood already at mean-field level, the length scale of the observed microphase separated steady states depends on the noise level and is therefore not a mean-field property. Moreover these steady states show manifest TRS breaking at least at intermediate, if not larger, length scales. The peculiar interplay of noise and activity in this model suggests that a closer investigation of its critical phenomena, using the tools of RG, is merited. With this motivation, we next study the various critical points of the AMB+ model by employing a perturbative dynamical RG to one-loop order. 

%%%%%%%%%%%%%%%%%%%%%%%%%%%%%%%%%%%%%%%%%%%%%%%%%%%%%%%%%%%%%%%%
\section{One-loop RG}\label{sec:RG-intro}
Close to the Gaussian fixed point we can assume $K$ and $D$ are fixed under the RG flow~\cite{Tauber2014}. Standard dimensional analysis then shows that $u$ is irrelevant for $d>4$, while $\lambda, \zeta$ and $\nu$ are irrelevant for $d>2$. This simple argument seems to lead to the conclusion that, in the physically relevant dimensions $d=2$ and $d=3$, any type of phase separation in AMB+ should be ruled by the Wilson-Fisher fixed point of passive Model B, and that activity should not have any impact on the physics involved.

However, dimensional analysis does not always lead to the correct conclusion. A famous example is in the Kardar-Parisi-Zhang equation where, although the nonlinearity is formally irrelevant above $d=2$, a strong coupling fixed point is present for $d\geq 2$. Even now it remains unclear whether any upper critical dimension exists for the KPZ equation~\cite{Tauber2014,canet2010nonperturbative}. 
Moreover, when $a=u=\nu=0$, AMB+ reduces to a surface growth model called cKPZ+~\cite{caballero2018strong}. This describes a conserved version of the KPZ dynamics with an additional nonlinearity given by the $\zeta$ term, alongside the traditional  KPZ nonlinearity described by $\lambda$. This model of roughening surfaces was only recently introduced and studied both with one-loop RG and numerical simulations~\cite{caballero2018strong}. Its RG flow closely resembles the one of KPZ, with a strong coupling fixed point present for $d\geq 2$. In addition, numerical simulations seem to support the presence of a strong coupling fixed point in $d=2$. 

This motivates the one-loop analysis that we perform below. We will conclude that,
while the active nonlinearities are formally irrelevant for any $d>2$, this irrelevance is reflected in the physical behavior only when the system undergoes bulk phase separation. In contrast, we argue that the transition to microphase separation is likely connected with a strong-coupling regime whose existence is predicted by our one-loop calculation even thought exploring its full character lies beyond our perturbative approach. This suggests that in the microphase-separation regime, active terms have a controlling influence on the large scale physics in dimensions $d=2,3$, as already
indicated at mean-field level~\cite{tjhung2018reverse}.

We first transform (\ref{eq:AMB+}) into Fourier space with wavevector $\bfq$ (of modulus $|\bfq |=q$) and frequency $\omega$:
\qq\label{eq:AMB+Fourier}
\phi(\hat{\bfq})&&=
\phi_0(\hat{\bfq})+\frac{G_0(q,\omega)}{2}\int_{\hat{\bfq}'}g(\bfq,\bfq')\phi(\hat{\bfq}')\phi(\hat{\bfq}-\hat{\bfq}')\nonumber\\
&&-u G_0(q,\omega)\int_{\hat{\bfq}', \hat{\bfq}''} \phi(\hat{\bfq}')\phi(\hat{\bfq}'')\phi(\hat{\bfq}-\hat{\bfq}'-\hat{\bfq}'')
\qqq
where $\hat{\bfq}=(\omega,\bfq)$, 
\qq \label{eq:phi0def}
\phi_0(\hat{\bfq})=\frac{G_0(q,\omega)}{q^2}\eta(\hat{\bfq})\,,
\qqq
the  bare propagator is 
\qq
G_0(q,\omega) = \frac{q^2}{ (-i\omega +aq^2+K q^4)}
\qqq
and $\eta$ is a Gaussian white noise with zero average and variance $\langle\eta(\hat{\bfq})\eta(\hat{\bfq}')\rangle = 2Dq^2(2\pi)^{d+1}\delta^{d+1}(\hat{\bfq}+\hat{\bfq}')$. 
\begin{figure}\center
	\begin{minipage}{0.2175\linewidth}\centering\vspace{-0.78cm}\includegraphics[page=1]{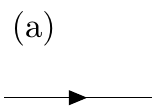}\end{minipage}
	\begin{minipage}{0.2175\linewidth}\centering\vspace{-0.78cm}\includegraphics[page=2]{diag_defs}\end{minipage}
	\begin{minipage}{0.2175\linewidth}\centering\includegraphics[page=3]{diag_defs}\end{minipage}
	\begin{minipage}{0.2175\linewidth}\centering\includegraphics[page=4]{diag_defs}\end{minipage}
	\caption{Diagrammatic notations used: a line denotes a zeroth-order field $\phi_0$ (a); the correlation function $C_0(q,\omega)$, is represented as a circle between two incoming lines (b). We have two vertices. The one corresponding to the $\lambda,\zeta,\nu$ non-linearities is represented in (c) and reads $(G_0(q,\omega)/2) \int_{\hat{\bfq}'}g(\bfq,\bfq')\phi(\hat{\bfq}')\phi(\hat{\bfq}-\hat{\bfq}')$, 
with $\hat{\bfq}$ the wavevector entering into the vertex. The vertex corresponding to $u$, represented in (d), is instead $-u G_0(\hat{\bfq})\int_{\hat{\bfq}', \hat{\bfq}''} \phi(\hat{\bfq}')\phi(\hat{\bfq}'')\phi(\hat{\bfq}-\hat{\bfq}'-\hat{\bfq}'')$.
	\label{fig:diagrams-notation}}
\end{figure}
We also denote the two-point correlation function of the linear theory ($u=\lambda=\zeta=\nu=0$) by 
\qq
C_0(\hat{\bfq},\hat{\bfq}')&=&(2\pi)^{d+1} C_0(q,\omega) \delta^{d+1}(\hat{\bfq}+\hat{\bfq}')\\
C_0(q,\omega) &=& \frac{2D}{q^2} G_0(q,\omega) G_0(-q,-\omega)\,.
\qqq
In (\ref{eq:AMB+Fourier}), the nonlinearities $\lambda$, $\zeta$ and $\nu$ enter via the function $g(\bfq,\bfq')$ that (after symmetrising, $\bfq'\leftrightarrow(\bfq-\bfq')$, without loss of generality) reads 
\qq\label{eq:vertex-g}
g(\bfq,\bfq') = 
2\lambda \bfq'\cdot(\bfq-\bfq')
+\nu q^2
-\frac{\zeta}{q^2}\left[
q'^2\bfq\cdot(\bfq-\bfq') +|\bfq-\bfq'|^2 \bfq\cdot \bfq'\right]\,.
\qqq
Diagrammatic notation is introduced following the rules in Fig. \ref{fig:diagrams-notation}.

\begin{figure}
	\begin{minipage}{0.2175\linewidth}\centering\vspace{-0.9cm}\includegraphics[page=1]{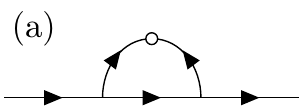}\end{minipage}
	\begin{minipage}{0.2175\linewidth}\centering\includegraphics[page=2]{diag_all}\end{minipage}
	\begin{minipage}{0.2175\linewidth}\centering\includegraphics[page=3]{diag_all}\end{minipage}
	\begin{minipage}{0.2175\linewidth}\centering\vspace{-0.3cm}\includegraphics[page=4]{diag_all}\end{minipage}
	
	\begin{minipage}{0.2175\linewidth}\centering\vspace{+0.45cm}\includegraphics[page=5]{diag_all}\end{minipage}
	\begin{minipage}{0.2175\linewidth}\centering\vspace{+0.4cm}\hspace*{+0.5cm}\includegraphics[page=6]{diag_all}\end{minipage}
	\begin{minipage}{0.2175\linewidth}\centering\includegraphics[page=7]{diag_all}\end{minipage}
	\begin{minipage}{0.2175\linewidth}\centering\vspace{+0.45cm}\includegraphics[page=8]{diag_all}\end{minipage}
	
	\caption{ All the diagrams at $1$-loop. Their contributions are discussed in the text.
	\label{fig:diagrams}}
\end{figure}

 We next apply dynamical RG with dimensional regularisation~\cite{Tauber2014}, using the small parameter $\epsilon=4-d$, meaning that we perform a perturbative expansion just below the upper critical dimension found by dimensional analysis ($d_c=4$). At $1$-loop, all the diagrams shown in Fig.~\ref{fig:diagrams} might contribute in principle. We give the explicit computation of the diagrams in ~\ref{fig:diagrams}(a,f) in \ref{app:1a} and \ref{app:1f}. All other diagrams can be computed using the same techniques so we leave out their explicit derivation. Here we summarise the results of these lengthy computations.

First, we observe that the diagrams in Fig.~\ref{fig:diagrams}(g) and \ref{fig:diagrams}(h) only involve the $\phi^4$ vertex, so their contribution is exactly the same as in passive $\phi^4$ theory (Model B) which can be found in the literature~\cite{hohenberg1977}. Second,  
the diagram in Fig.~\ref{fig:diagrams}(d) may in principle renormalize the noise strength $D$, but a closer look shows that its first nonvanishing order in $q$ is $q^4$, thus producing only higher-order terms in the noise which can be safely discarded close to $d=4$. Again, an explicit computation is not necessary. All other diagrams, given in Fig.~\ref{fig:diagrams}(a), \ref{fig:diagrams}(b), \ref{fig:diagrams}(c), \ref{fig:diagrams}(e), \ref{fig:diagrams}(f), have to be computed explicitly.

Second, we observe that, due to the presence of the $\lambda,\zeta,\nu$ terms, AMB+ is not symmetric under the $\phi\to-\phi$ transformation (in contrast to passive Model B). It is then not surprising that the diagrams in Fig.~\ref{fig:diagrams}(b),(c),(e),(f) produce a cubic nonlinearity which enters in the equation for $\p_t\phi$ as $\nabla^2(c\phi^2)$. We absorb this term by an additive shift of the field. This ensures that the critical density remains at $\phi = 0$ under the RG flow and is equivalent to demanding that $\phi$ is always defined relative to the critical density (at which the transition is second order). 
A similar procedure is standard in the study of the liquid-vapor critical point where there is in general no symmetry between positive and negative order parameters. Despite this, that system lies in the symmetric $\phi^4$ (Ising) universality class, because the critical point is precisely the one at which cubic terms vanish and symmetry is restored.
We discuss below how this cubic nonlinearity does not change the flow equations around the critical point.

Third, upon integrating over a thin momentum shell $q\in[\Lambda/(1+db),\Lambda]$, with $\Lambda$ the ultraviolet cutoff, we thereby obtain the following intermediate (subscript $I$) values of the coupling constants:
\qq
a_I &= a+3K\bar u\Omega_d\Lambda^{d-2}db+\frac{K\bar\nu}{2d}\left[(d-2)\bar\zeta+2d\bar\lambda\right]\Omega_d\Lambda^ddb\label{eq:int1}\\
D_I &= D\\
K_I &= K\left(1-M\Omega_d\Lambda^{d-2}db\right)\\
u_I &= u\left(1-9\bar u\Omega_d\Lambda^{d-4}db\right)\\
\nu_I &= \nu+\frac{K^{3/2}}{D^{1/2}}\left(T_{\bar \nu}\Lambda^{d-2}+B_{2,\bar \nu}\Lambda^{d-4}+B_{1}\Lambda^{d-4}\right)\Omega_ddb\\
\lambda_I &= \lambda+\frac{K^{3/2}}{2D^{1/2}}\left(T_{\bar \lambda}\Lambda^{d-2}+B_{2,\bar \lambda}\Lambda^{d-4}\right)\Omega_ddb\\
\zeta_I &=\zeta-\frac{K^{3/2}}{D^{1/2}}T_{\bar \zeta}\Lambda^{d-2}\Omega_ddb
\qqq
where we have rewritten all non trivial contributions in terms of the reduced couplings $\bar\lambda^2=\lambda^2DK^{-3/2}$, $\bar\zeta^2=\zeta^2DK^{-3/2}$, $\bar\nu = \nu^2DK^{-3/2}$, $\bar u = uDK^{-2}$, $\bar a = a/K$, $\Omega_d = S_d/(2\pi)^d$, with $S_d$ the surface of the $d$-dimensional sphere. In the above equations
\qq
M = \frac{1}{2d(d+2)} \Big[&(d-2) (2 d+1) \bar\zeta ^2+d \bar\zeta  ((4-d) \bar\nu +4 (d+2) \bar\lambda )\nonumber\\
& -(d+2) \left(2d \bar\lambda  \nu -d \bar\nu ^2+4\bar\lambda ^2\right)\Big] \label{eq:full-flow-M}
\qqq
\qq
T_{\bar \nu} = \frac{\bar\nu}{d(d+2)}\Big[&(d-2) (2 d+1) \bar\zeta ^2+d \bar\zeta  ((4-d) \bar\nu +4 (d+2) \bar\lambda )\nonumber\\
&-(d+2) \left(2d \bar\lambda  \bar\nu -d \bar\nu ^2+4\bar\lambda ^2\right)\Big]
\qqq
\qq
T_{\bar \lambda} = \frac{\bar\nu}{4d(d+2)}\Big[& -2 (d-2) (7 d+4) \bar\zeta ^2 -4 (d+2) \bar\lambda  (2(d-2) \bar\lambda -3 d \bar\nu )\nonumber\\
&-4 \bar\zeta  (2(d (4 d+5)-10) \bar\lambda -(d-2) d \bar\nu )\Big]
\qqq
\qq
T_{\bar \zeta} &=& \frac{2\bar\nu\bar\zeta\left[4(d-3)\bar\zeta-8(1+d)\bar\lambda-d(6+d)\bar\nu\right]}{4d(d+2)}\\
B_{1} &=& \frac{3u\left[2(4-d)\bar\zeta-(2+d)(4\bar\lambda+d\bar\nu)\right]}{d(d+2)}\\
B_{2,\bar \nu} &=& \frac{3 u  \left[2 (d-1) \bar\zeta -(d-2) \bar\nu \right]}{d}\\
B_{2,\bar \lambda} &=& -\frac{6 u \left[2 (d-1) \bar\zeta -(d-2) \bar\nu \right]}{d}.\label{eq:intlast}
\qqq

The fourth step to obtain the RG flow equations is to rescale the equations of motion, in which the parameters are now given by the intermediate values, to restore the original momentum. This means applying the transformation $q\to bq$, with $b = 1+db$, leaving two free exponents for the time frequency and the field: $\omega\to b^z\omega$ and $\phi\to b^{-\chi}\phi$. In Model B, the transition between uniform and fullly phase separated states, governed by the Wilson-Fisher fixed point, can be accessed by asking that $K$ and $D$ are kept fixed under the RG flow. We make the same choice here of fixing $K$ and $D$ under the RG flow, so that in the limit of no activity we recover the equilibrium results. Reabsorbing all the scaling factors into the couplings, with primes denoting their new values, we obtain:
\qq
a' &= b^{z-2}a_I\label{eq:disc_change1}\\
K' &= b^{z-4}K_I\\
u' &= b^{z-2+2\chi}u_I\\
D' &= b^{z-2-d-2\chi}D_I\\
(\nu', \lambda', \zeta') &= b^{z+\chi-4}(\nu_I, \lambda_I, \zeta_I)\label{eq:disc_change2}
\qqq
where the intermediate values are given in the previous section.

Lastly we let $db$ become infinitesimal and consider the resulting parameter increments $a'=a+da$, etc.. Expanding both sides of (\ref{eq:disc_change1}-\ref{eq:disc_change2}) and writing the result in terms of reduced couplings as defined above (which removes the dependence of the flow equations on $z$ and $\chi$) we finally obtain:
\qq
\frac{d\bar a}{db}  &=&\label{eq:full-flow1} 2\bar a+3\bar u\Omega_d\Lambda^{d-2}+\frac{\bar\nu}{2d}\left[(d-2)\bar\zeta+2d\bar\lambda\right]\Omega_d\Lambda^d+\bar aM\Omega_d\Lambda^{d-2}\\
\frac{d\bar u}{db} &=& \bar u(4-d) - 9\bar u^2\Omega_d\Lambda^{d-4}+2\bar uM\Omega_d\Lambda^{d-2}\label{eq:full-flow2}\\
\frac{d\bar \nu}{db} &=& \bar \nu\left(\frac{2-d}{2}+\frac{3}{2}M\Omega_d\Lambda^{d-2}\right)+\left(T_{\bar \nu}\Lambda^{d-2}+B_{2,\bar \nu}\Lambda^{d-4}+B_{1}\Lambda^{d-4}\right)\Omega_d\label{eq:full-flow3}\\
\frac{d\bar \lambda}{db}&=&\bar \lambda\left(\frac{2-d}{2}+\frac{3}{2}M\Omega_d\Lambda^{d-2}\right)+\frac{1}{2}\left(T_{\bar \lambda}\Lambda^{d-2}+B_{2,\bar \lambda}\Lambda^{d-4}\right)\Omega_d\label{eq:full-flow4}\\
\frac{d\bar \zeta}{db} &=& \bar \zeta\left(\frac{2-d}{2}+\frac{3}{2}M\Omega_d\Lambda^{d-2}\right)-T_{\bar \zeta}\Omega_d\Lambda^{d-2}\label{eq:full-flow5}
\qqq

In the steps needed to go from the intermediate values to the final flow equations, we have so far ignored the additive renormalization of the $\phi$ field that was previously introduced to eliminate a term $cq^2\phi^2$ generated in the equation of motion. However, this does not change the flow equations: in the expression $\nabla^2(a\phi+c\phi^2+u\phi^3)$, the field shift required to eliminate the quadratic term is $\phi\to \phi+c/(3u)$ and the only other parameter modified is $a\to a-5c^2/(3u)$. Since $c$ is of order $db$, the resulting shift in $a$ is quadratic in $db$ and therefore does not enter the flow equations.

The RG flow equations in (\ref{eq:full-flow1}-\ref{eq:full-flow5}) represent the main technical results of this paper and they will be further analysed in what follows. 
Two observations are due immediately. First, and most obviously, the flow in (\ref{eq:full-flow1}-\ref{eq:full-flow5}) reduces to that of Model B when $\bar\lambda=\bar\zeta=\bar\nu=0$. Second, the flow respects time-reversal symmetry if we start from any equilibrium model as our initial condition: if at bare level we have that $\bar\zeta=0$ and $2\bar\lambda=-\bar\nu$, both conditions are maintained along the flow. 

%%%%%%%%%%%%%%%%%%%%%%%%%%%%%%%%%%%%%%%%%%%%%%%%%%%%%%%%%%%%%%%%
\section{RG flow for $\bar\nu=0$: strong coupling}\label{sec:nu0}
The RG flow in (\ref{eq:full-flow1}-\ref{eq:full-flow5}) is rather complex. In order to get physical insight, we first analyse it while assuming that $\bar\nu=0$, not only at bare level but all along the flow. Note however that even if $\bar\nu=0$ at bare level, an inspection of (\ref{eq:full-flow3}) reveals that it acquires a non-zero value as the flow proceeds. 
Barring some physical mechanism that would prevent this from happening (and we have not identified one), the analysis of this section represents an approximation involving a continuous projection of the differential flow onto the $\bar\nu = 0$ subspace.  However, the approximate (or `projected') flow that results is much more clearly understood than the full one, and moreover it produces results that are qualitatively very similar to the full flow as will be shown in Sec.~\ref{sec:full-RG-flow}. 

\begin{figure*}\center
	\includegraphics[scale=0.4]{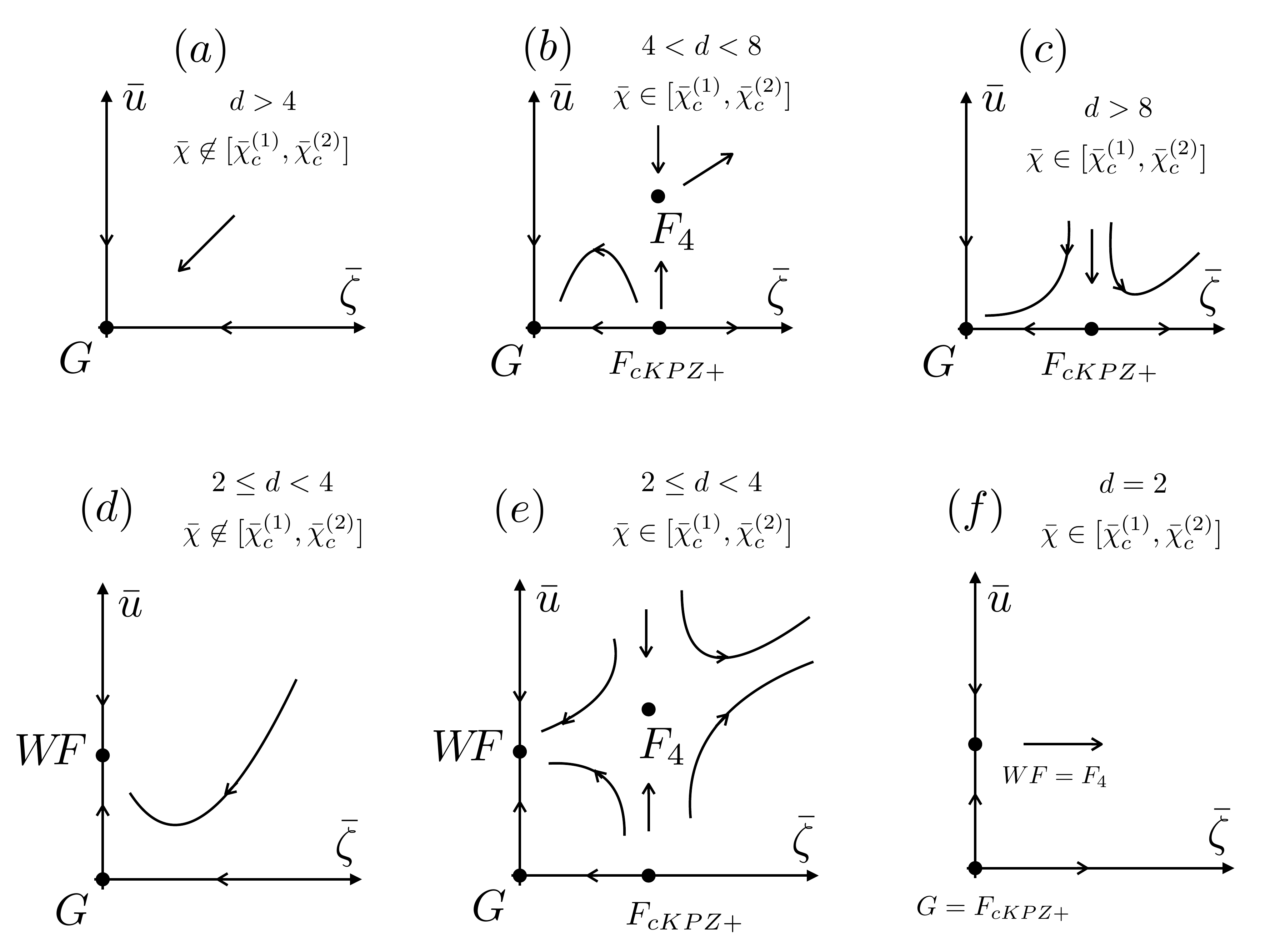}
	\caption{$1$-loop RG flow of AMB+ for $d\geq2$ and $\nu=0$. With respect to Model B, where only the Gaussian (G) and Wilson-Fisher (WF) fixed points are present, two new ones arise ($F_{cKPZ+}$ at $\bar u=0$ and $F_4$ at $\bar u, \bar\zeta\neq 0$) when $\bar\chi\in [\bar\chi_c^{(1)},\bar\chi_c^{(2)}]$. $F_{cKPZ+}$ is fully repulsive while $F_4$ is attractive in the $\bar u$ direction but repulsive in the $\bar \zeta$ one. The WF fixed point is locally attractive in all directions for $2<d<4$. 
\label{fig:RGflow}}
\end{figure*}

Setting $\bar\nu=0$, the topology of the projected flow can be obtained analytically. Indeed, in this case, the flow is most conveniently written in terms of $\bar \zeta$ and of $\bar \chi=\bar \lambda/\bar\zeta$. We then find that the latter does not flow at all and, in this representation, we obtain
\qq
\frac{d\bar a}{db}&=2\bar a + 3\bar u\Omega_d\Lambda^{d-2}+\bar aN\Omega_d\Lambda^{d-2},\\
\frac{d\bar u}{db}&=\bar u\left(4-d-9\bar u\Omega_d\Lambda^{d-4}+2N\Omega_d\Lambda^{d-2}\right),\\
\frac{d\bar\zeta}{db}&=\bar\zeta\left(\frac{2-d}{2}+\frac{3}{2}N\Omega_d\Lambda^{d-2}\right),\label{eq:flow-l-z}\\
\frac{d\bar\chi}{db}&=0\,,
\qqq
where 
\qq
N &\equiv M(\bar\lambda=\bar\chi \bar \zeta ,\bar\nu=0)\\
&=- \frac{\bar\zeta^2}{2d(2+d)}\left[-4 (d+2) \chi ^2+4 d (d+2) \chi +d (2 d-3)-2\right]\,,
\qqq
with $M$ given in (\ref{eq:full-flow-M}).
The projected RG flow is graphically represented in Fig.~\ref{fig:RGflow} for different values of $d$ and $\bar\chi$.

%flow
%conclusions

We now make two observations.
First,  when $\lambda=\zeta=0$, we recover the RG flow of Model B, with the Gaussian fixed point $u=0$ attractive for $d>4$ and repulsive for $2\leq d<4$. In the latter case, the Wilson-Fisher (WF) fixed point becomes attractive. 
Second, when the $\phi^4$ nonlinearity is absent ($u=0$), AMB+ at $a = a_c = 0$ reduces to the surface growth model cKPZ+, describing a particular type of roughening surface as described in~\cite{caballero2018strong}. In the latter work it was shown at one loop that in a specified region of $(\lambda,\zeta)$ parameters, 
the RG flow of cKPZ+ leads to strong coupling for any $d\geq2$, even though naive dimensional analysis would lead to the conclusion that all nonlinearities are irrelevant. This shows up in the projected RG flow via the fact that the Gaussian fixed point is unstable in $d=2$ for a certain parameter range of $\bar\lambda, \bar\zeta$. For $d>2$, the Gaussian fixed point is then stable, but a second fixed point $F_{cKPZ+}$, with $(u=0,\zeta\neq 0)$, is created, beyond which the flow diverges to infinity. The $F_{cKPZ+}$ fixed point is represented in Fig.~\ref{fig:RGflow} in cases where it exists.

For a more precise discussion, taking advantage of the symmetry $(\lambda,\zeta,\phi)\to -(\lambda,\zeta,\phi)$, we can restrict ourselves to the case $\bar\zeta>0$. First, the Gaussian fixed point is locally attractive for $d>4$ and the WF fixed point is locally attractive when $2< d<4$, see Fig.~\ref{fig:RGflow}. 
Moreover, the WF fixed point is globally attractive when
 $2\leq d\leq 4$, and $\bar\chi\not\in\left[\bar\chi_c^{(1)},\bar\chi_c^{(2)}\right]$, where
\qq
\bar\chi_c^{(1)}&=& d-\sqrt{d (d+2)+\frac{12}{d+2}-7}\\
\bar\chi_c^{(2)}&=& d+\sqrt{d (d+2)+\frac{12}{d+2}-7}\,.
\qqq
Up to this point, our results could have been expected a-priori from dimensional analysis. The most interesting fact is, however, that when $2\leq d< 4$ and $\bar\chi\in\left[\bar\chi_c^{(1)},\bar\chi_c^{(2)}\right]$, while the WF fixed point is locally stable (marginally in $d=2$), two new fixed point of the projected RG flow appear. One of them is that of the cKPZ+ equation $F_{cKPZ+}$ already discussed above, while the second appears at non-vanishing $\bar u$ and $\bar\zeta$ and will be called $F_4$ hereafter.
The two fixed points $F_{cKPZ+}$ and $F_4$ converge to, respectively, the Gaussian and the WF fixed point in the limit $d\to 2^+$. Moreover, while $F_{cKPZ+}$ is unstable both along the $\bar u$ and the $\bar \zeta$ directions, $F_4$ is unstable along the $\bar \zeta$ but stable along $\bar u$.
The non-equilibrium fixed point $F_4$ that we found in this study represents a modification of the WF fixed point; we can consider $\bar\zeta$ and $\bar\chi$ as control parameters and thus $F_4$ governs the transition between critical bulk phase separation, and a new behavior arising when the one-loop RG flow is towards strong coupling. All we know about the AMB+ model at both mean-field and numerical levels~\cite{tjhung2018reverse} suggests that this new behavior represents microphase separation. If so, $F_4$ should be closely related to the Lifshitz point of an equilibrium system (although we will see in the next section that it is not identical).

Summarising, the above analytic results for the projected RG flow imply that the critical point separating a homogeneous phase from bulk phase separation falls in the same universality class as Model B, and is governed by the WF fixed point. In contrast, the critical transition from bulk to microphase separation is governed by a new $F_4$ fixed point. The critical exponents at $F_4$ can be found formally at $1$-loop by simultaneously expanding in $4-d$ and $d-2$:
\qq
z&=&4+\frac{d-2}{3}\\
\chi&=&\frac{2-d}{3}\\
\nu^{-1}&=&2+c_4+\frac{d-2}{3}.
\qqq
Here $z$ and $\chi$ are calculated by fixing $K$ and $D$ under the projected RG flow, and  $\nu^{-1}$ (not to be confused with the coupling $\nu$), is the exponent associated with the divergence of the correlation length, calculated by linearizing the flow of $\bar a$ close to the transition point. Finally, $c_4=-(4-d)/3+o((4-d)^2)$ is the correction to the correlation length exponent coming from the passive $\phi^4$ theory. (Recall that there are no similar corrections to $z$ or $\chi$ at one-loop level~\cite{onuki2002phase}). The rest of the contributions to the exponents are calculated to first order in the $d-2$ expansion. 

We note that the critical exponents given above should be trusted only qualitatively. Indeed, the structure of the flow summarised in Fig.~\ref{fig:RGflow} implies that there is no regime where $F_4$ can be accessed perturbatively: for $d\to4^-$ dimensions the value of $\bar u$ at $F_4$ is small but the value of $\bar \zeta$ is not. Instead, $\bar \zeta$ becomes small in the limit $d\to2^+$ ($F_4$ approaches the WF fixed point) but $\bar u$ is not. Put differently, this limitation of the one-loop calculation is reflected in the requirement to expand simultaneously in $4-d$ and $d-2$ which cannot, of course, both be small at once.

As mentioned already above, our study of the projected flow with $\bar\nu=0$ has the advantage of giving a simple result that can be treated analytically. Importantly, most conclusions drawn above from doing this still hold qualitatively for the full flow in which $\bar\nu\neq 0$. This is the topic of the next section.

%%%%%%%%%%%%%%%%%%%%%%%%%%%%%%%%%%%%%%%%%%%%%%%%%%%%%%%%%%%%%%%%
\section{Complete RG flow}\label{sec:full-RG-flow}
\begin{figure}
\center \includegraphics[scale=0.5]{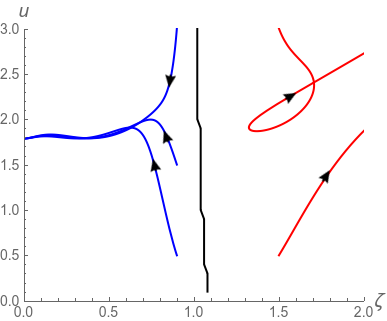}
	\caption{Numerical integration of the full RG flow in (\ref{eq:full-flow1}-\ref{eq:full-flow5}) obtained for $5$ different initial conditions, all of them with bare values $\bar\lambda=3.24, \bar\nu=0$ and dimension $d=2.5$.
	The flow is represented in the ($\bar\zeta$-$\bar u$)-plane. As it was the case in Fig. \ref{fig:RGflow}, a separatrix (black, nearly vertical line) between two different regimes appears: when $\bar\zeta$ is small (blue lines), the flow converges back to WF fixed point (we checked that also $\bar\lambda$ and $\bar\nu$ are flowing to zero). On the right of the separatrix (red lines), the flow diverges to infinity (strong coupling).} 
	\label{fig:full-RGflow}
\end{figure}
We now turn to analyse the full RG flow given in eq. (\ref{eq:full-flow1}-\ref{eq:full-flow5}). Technically, this is much harder because the flow is no longer radial in the $(\bar\lambda,\bar\zeta)$ plane, making $\bar\chi$ not conserved along the flow. We thus restrict ourselves to a numerical analysis of the flow close to the WF fixed point when $2\leq d<4$. 

We employ the Newton method to find the fixed points of the flow. Restricting to $\bar\zeta>0$, we found two fixed points  that merge into the WF one when $d \to 2^+$; one of them is the generalization of $F_4$, while the other one, which we will refer to as $F_{\textrm{eq}}$, represents an equilibrium fixed point. This satisfies the condition $2\lambda=-\nu$, $\zeta=0$, and thus the dynamics at this fixed point admits a free energy functional. 
The flow equations have several other fixed points that we do not explore here, as they do not connect to the Gaussian or WF fixed points as $d\to2^+$. Accordingly these additional fixed points are most likely artefacts of the 1-loop computation.

The stability of $F_4$ and $F_{\textrm{eq}}$ can be studied numerically. In dimensions $2<d<4$, the WF point is still locally attractive; $F_4$, as in Section \ref{sec:nu0}, has one unstable direction. Instead, $F_{\textrm{eq}}$ has two unstable directions, one of which is along the equilibrium subspace. 
When $d\to 2^+$, the collapse of $F_4$ and of $F_{\textrm{eq}}$ onto WF make the WF fixed point marginally unstable along two directions. 
This means that the linearization of the flow in $d=2$ around the WF fixed point has two zero eigenvalues but the flow is non-linearly unstable. One of these two directions is along the equilibrium subspace.

Note that the presence of an unstable fixed point in the equilibrium subspace might have been expected. This is because this subspace effectively describes models with a square gradient coefficient of the form $K(\phi) = K+2K_1\phi$ so that the system is unstable at initial densities $\phi_0$ (either positive or negative) of sufficient magnitude. Although the theory is set up close to $\phi = 0$, we have already noted the generation of a cubic term which is removed by an additive renormalization of $\phi$. In this process the value of $K$ should also change; depending on the sign of $K_1$ it may move towards negative or positive values. Should negative $K$ ultimately result, the system becomes unstable;  in an equilibrium context stability must be restored by adding gradient terms in the free energy. 
Such an equilibrium model gives finite-wavelength ordering (with wavelength fixed by the free-energy structure) and describes microphase separated states ~\cite{gompper1994amphiphilic}. The point of first instability $K = 0$ is then a new critical point, the Lifshitz point, at which the wavelength of this ordering diverges. We offer this as the interpretation of $F_{\textrm{eq}}$.

Notably, however, the microphase separated states in AMB+ are found to be stable (at least in the case of $\bar\nu=0$) without the need to introduce higher order derivatives; moreover, the length scale of these states is found numerically to be controlled by the noise level~\cite{tjhung2018reverse}. We therefore consider it unlikely that the transition from bulk phase separation to microphase separation arising in AMB+ lies in the same universality class as the equilibrium Lifshitz point. This is supported by the fact that our RG flow carries any active system (one whose initial parameters are not on the equilibrium subspace), not to $F_{\textrm{eq}}$, but instead to $F_4$. Thus $F_4$ is a strong candidate for a {\em nonequilibrium counterpart} of the Lifshitz point. Apart from when they merge at $d = 2+$, these remain distinct fixed points and so there can be no expectation that they share a universality class. Note also that, for an equilibrium system beyond the Lifshitz point, fluctuations drive the microphase separation transition first order: the system always jumps direct from a uniform phase to one where the ordering is of finite amplitude. Therefore there are no critical exponents to be calculated in this regime; the first order behavior is properly captured by a self-consistent calculation~\cite{brazovskii1975phase,fredrickson1987fluctuation} and not an RG one. It is not clear at present whether the flow to strong coupling (arising beyond the separatrix on which both $F_{\textrm{eq}}$ and $F_4$ reside) is itself a signature of this first order transition. Our RG approach cannot therefore shed light on whether activity-induced microphase separation is also generically first order. 

Finally, our numerical integration of the full flow equations (\ref{fig:full-RGflow}) gives qualitatively similar results to those shown in Fig.~\ref{fig:RGflow}(e) for the projected flow in which $\nu$ is held fixed at zero.
In Fig.~\ref{fig:full-RGflow} we show the results of such numerical integration obtained using five different initial conditions and projecting the final results on the subspace $(\bar\zeta,\bar u)$. (Unlike the `projection' of the previous section this is not an approximation, but simply a way of representing the results on the printed page.) 
Note that in this projected representation the full flow can self-intersect as happens in one of the trajectories shown. Of course this does not happen if the flow trajectories are instead represented in the full $(u,\lambda,\zeta,\nu)$-space.
Two of our initial conditions (represented in red) are chosen in the region where the flow does not converge to the WF fixed point, while three of them (represented in blue) are in the region where the flow converges to WF fixed point. The aforementioned unstable fixed point belongs to the black line, which represents the critical surface. 

We conclude from various numerical results discussed above that the full RG flow of AMB+, at least in the proximity of the WF fixed point, is qualitatively similar to the one obtained imposing $\nu=0$ as we did in Section \ref{sec:nu0}. If so, the physical conclusions drawn there still hold.

%%%%%%%%%%%%%%%%%%%%%%%%%%%%%%%%%%%%%%%%%%%%%%%%%%%%%%%%%%%%%%%%
\section{Conclusions}\label{sec:conclusions}

Many works concerning phase separation in scalar active matter have either directly relied upon the idea that an effective equilibrium picture emerges at large scales, or used this possibility to motivate specific approximation schemes~
\cite{Tailleur:08,Speck2014PRL,Farage:2015:PRE,Maggi:15,fodor2016far,szamel2016theory,Brady:15}. 
The value of this approximation depends on where the parameters of the system lie in relation to the dynamical phase diagram that governs steady-state behavior. 

For the model studied in this paper (Active Model B+, or AMB+), as well as a first-order coexistence of liquid/vapor type ending in a critical point, it was shown in~\cite{tjhung2018reverse} that a qualitatively different phenomenology also arises, involving microphase separation instead of bulk phase separation, at large activity levels.  (This microphase separation regime is absent in the subspace representing a simpler model, AMB, that does not include all the leading-order TRS-breaking contributions to the current~\cite{wittkowski2014scalar}.) The resulting phase transition between bulk and microphase separation was studied analytically at the level of mean-field kinetics, as should hold far from any critical points, in Ref.~\cite{tjhung2018reverse}. This study showed that the transition is due to the reversal, within specified parameter ranges, of the Ostwald mechanism. The latter provides the diffusive pathway to full bulk phase separation in equilibrium fluids. Notably, this form of active microphase separation arises despite the absence, within the passive sector of the model, of any free-energy terms that would directly promote ordering at a finite length scale. Such terms would, in equilibrium, allow the critical point for bulk phase separation to connect continuously to one for microphase separation via a Lifshitz point, at which the wavelength of the steady-state density pattern diverges smoothly~\cite{gompper1994amphiphilic}. 

In this paper we have presented the first Renormalization Group study of critical phenomena within Active Model B+. Our main results are twofold. First, we showed that the critical point describing the transition from a uniform phase to bulk phase separation lies in the same equilibrium universality class as for passive Model B: the low noise phase is controlled by standard WF fixed point, and time-reversal-symmetry breaking terms flow to zero upon coarse graining. This is exactly what one would have expected from the dimensional analysis of the couplings that can be added to Model B to break time-reversal symmetry at leading order in $\nabla,\phi$. Notably though, although this implies that exponents for the divergence correlation lengths, relaxation times {\em etc.} coincide with those of passive Model B, there could still be nontrivial irreversibility of the active model at this critical point -- codified, for instance, in a nontrivial scaling of the entropy production. Building on ideas developed in~\cite{Nardini2017}, we will explore this feature elsewhere~\cite{caballeroEntropy}.

Unexpectedly, however, we find that the critical phase point separating bulk phase separation from microphase separation (which would in equilibrium systems be the Lifshitz point) appears to be ruled by a new fixed point $F_4$ of the RG flow, which is repulsive along a single direction, described by a linear combination of the active couplings. This is at first sight surprising because standard dimensional analysis predicts that {\em all} active terms in AMB+ should be be irrelevant near the Wilson-Fisher point and thus flow to zero upon coarse graining. Instead the fixed point $F_4$ lies on the separatrix between systems that flow back towards the WF fixed point and those in which the active parameters flow towards strong coupling. Within our perturbative approach, it is not possible to identify with certainty the physics of the strong coupling regime but it is natural to assume that this corresponds to microphase separation. Notably, $F_4$ is distinct from $F_{\textrm{eq}}$ which also lies on the separatrix but governs the subspace of equilibrium models and is interpreted as a Lifshitz point. 
We therefore believe that our $F_4$ fixed point represents a counterpart of the Lifshitz point, occupying a new nonequilibrium universality class. It is possible that finite-wavelength microphase separation (lying beyond the Lifshitz point) could also be different for active and passive systems, although this cannot be established perturbatively. A hint that this is indeed the case is given by the fact that, as shown numerically in~\cite{tjhung2018reverse}, this finite length-scale depends on the noise amplitude in the active case, while it is selected at deterministic level in the equilibrium one.

As always happens when a perturbative calculation points to nonperturbative physics, our predictions, in particular those concerning the critical exponents at the $F_4$ transition, cannot be trusted quantitatively.  Indeed, within our one-loop analysis, we can find no upper critical dimension beyond which all non-linear couplings are small at $F_4$ (which is why we introduced a simultaneous expansion about $2$ and $4$ dimensions to study this fixed point). Further studies are clearly needed to characterize quantitatively the new non-equilibrium universality classes ruling microphase separation in active systems. In this direction, the application of non-perturbative RG~\cite{delamotte2012introduction,berges2002non} could shed further light on the surprisingly complex phase behavior of AMB+.
Fluctuations are also important deep within the microphase-separated phase. Indeed, as shown numerically in  \cite{AMBplus}, the finite length-scale selected here is strongly noise dependent and it remains an open question how to characterize it theoretically.

Finally, we note that a recent computational work~\cite{siebert2017critical} investigated a critical point for phase separation of a many-body active particle system. (The latter comprised repulsive Active Brownian Particles (ABPs), one of the systems for which AMB+ ought to serve as a good continuum description~\cite{stenhammar2013continuum}.)
The authors gave evidence that the static critical exponents differ from those of liquid-gas phase separation; if true this would place the system outside the passive Model B universality class, contrary to our predictions concerning the character of bulk phase separation in active materials. One possible explanation is that the system of~\cite{siebert2017critical} actually lies close to a microphase separation (which, in the context of ABPs, correponds to vapor droplets in a liquid continuum or ``bubbly phase separation''~\cite{tjhung2018reverse}). If so, the reported observations in~\cite{siebert2017critical} could be governed by either the $F_4$ fixed point, or the strong coupling regime that lies beyond. However, this explanation remains speculative:  we are not aware of any particle-based simulation studies that unambiguously probe the critical region for active microphase separation. We hope our work will stimulate such studies, as well as a closer investigation of the critical behavior of continuum models of activity such as AMB+. 

%%%%%%%%%%%%%%%%%%%%%%%%%%%%%%%%%%%%%%%%%%%%%%%%%%%%%%%
%\textit{Acknowledgements.} 

\section*{Acknowledgments}
We acknowledge F. van Wijland for several useful discussions.
FC is funded by EPSRC DTP IDS studentship, project number $1781654$.
CN acknowledges the hospitality provided by DAMTP, University of Cambridge while part of this work was being done. 
CN acknowledges the support of an Aide Investissements d'Avenir du LabEx PALM (ANR-10-LABX-0039-PALM). Work funded in part by the European Resarch Council under the EU's Horizon 2020 Programme, grant number 760769. MEC is funded by the Royal Society. 

\appendix
%%%%%%%%%%%%%%%%%%%%%%%%%%%%%%%%%%%%%%%%%%%%%%%%%%%%%%%
%\textit{Acknowledgements.} 
\section{Computation of the diagram in Fig. \ref{fig:diagrams}(a)}\label{app:1a}

	Following the diagram rules defined in the main text, the diagram reads:
	\qq
	\vcenter{\hbox{\vspace*{0.4cm}\includegraphics[page=1]{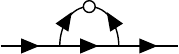}}}=4G_0({\hat{\bfq}})\int_{\hat{\bfq},\hat{\bfq}_I}&\frac{1}{2}\Bigg[&2\lambda \bfq_I\cdot(\bfq-\bfq_I)-\zeta\frac{q_I^2\bfq\cdot(\bfq-\bfq_I)}{q^2}\nonumber\\
	&&-\zeta\frac{|\bfq-\bfq_I|^2\bfq\cdot \bfq_I}{q^2}+\nu q^2\Bigg]\times\nonumber\\
	&\frac{1}{2}\Bigg[&-2\lambda \bfq\cdot \bfq_I+\zeta\frac{q^2\bfq_I\cdot(\bfq-\bfq_I)}{|\bfq-\bfq_I|^2}\nonumber\\
	&&-\zeta\frac{q_I^2\bfq\cdot(\bfq-\bfq_I)}{|\bfq-\bfq_I|^2}+\nu |\bfq-\bfq_I|^2\Bigg]\times\nonumber\\
	&&G_0(\hat{\bfq}-\hat{\bfq}_I)C_0(\hat{\bfq}_I)\phi(\hat{\bfq}),
	\qqq
	where the prefactor is the symmetry factor, and the two parenthesis are the vertices. We see that the diagram is of course proportional to $\phi$ (to $\phi_0$ once we truncate the iterative expansion, where $\phi_0$ is defined in equation \ref{eq:phi0def}), so it will only produce contributions to the linear terms in the equation of motion: $\phi,\nabla^2\phi,\nabla^4\phi,..$.
	
We then perform a small wavelength expansion. Because this is done on external wavelengths (since internal momenta have high wavelengths that are integrated out), it means performing a Taylor expansion on $q$. It is useful to rewrite $\bfq\cdot \bfq_I=qq_I\cos\theta$. After doing this expansion in both the vertex function and the zero order propagator and correlator (note that we drop the incoming propagator $G_0$ since we divide by it to count the contributions to each coupling, and we also do not consider the external fields that multiply everything, this is just the loop), we get:
	\qq
	&\hbox{\vspace*{0.4cm}\includegraphics[page=1]{app_a}}=-\frac{D \nu  (-\zeta  \cos (2 \theta )+2\lambda )}{2 K ^2}+\nonumber\\
	&\frac{qD}{2 K ^2 q_I} \cos (\theta ) (-\zeta  \cos (2 \theta ) (-\zeta +2\lambda +2 \nu )-\zeta  (2\lambda +\nu )+2\lambda  (2\lambda +3 \nu ))+\nonumber\\
	&\frac{q^2D}{2 K ^2 q_I^2} (-\cos ^2(\theta ) \left(7 \zeta ^2-6 \zeta  \nu +4\lambda ^2\right)+2 \zeta ^2+4 \zeta  (\zeta -\nu ) \cos ^4(\theta )\nonumber\\
	&-\zeta  (\nu -4 \lambda )+\nu  (\nu -2\lambda )) + o(q^3).
	\qqq
	The first term is the zeroth order term, contributing to the mass term $a$: it is proportional to $\nu$, as it should be, since for $\nu=0$  an additional symmetry prevents this kind of term from being generated~\cite{caballero2018strong,Janssen1996}. The second term is odd in $q$, and therefore in $\cos\theta$, so it will vanish once we finish the loop integral by integrating all angles. The last term is quadratic in $q$ and so it contributes to $K$. This gives the expression for $M$ we reported in eq. (\ref{eq:full-flow-M}). To get the final expression we simply need to integrate all angles, here the only one appearing in the integral is $\theta$ which varies between $0$ and $\pi$.
	
	The integrals have to be done in continuous dimension, and there are two non trivial ones: the integral of $\cos^2\theta$ and $\cos^4\theta$. The first one can be done seeing how $\cos^2\theta=u_{x_d}$, where $u_{x_d}$ is a unit vector in the direction of the last axis in the coordinate system, so the integral will be:
	\qq
	\int_{\textrm{sphere}}\cos^2\theta&=\int u_{x_d}\\
	&=\frac{1}{d}\int\left(u_{x_1}+u_{x_2}+\ldots+u_{x_d}\right)
	=\frac{1}{d}\int1
	=\frac{1}{d}\frac{S_d}{(2\pi)^d},\nonumber
	\qqq
	where $S_d$ is the surface of the sphere and the $(2\pi)^d$ comes from Fourier transform of the equations of motion. The step between the second and third integrals can be done due to isotropy of the space.

	The integral of $\cos^4\theta$ can be done in a similar fashion, writing $\cos^4\theta=u_{x_d}^2$ and rewriting it as a generic isotropic tensor, the result being
	\[ \int_{\textrm{sphere}}\cos^4\theta=\frac{3}{d(d+2)}\frac{S_d}{(2\pi)^d}. \]
	
	Using this in the expression of the second order term of the diagram and simplifying, we get:
	\qq
	\vcenter{\hbox{\vspace*{0.4cm}\includegraphics[page=1]{app_a}}}\,{}_{2}=
	\frac{q^2 K }{2 d (d+2)}
	M\int_{\Lambda/b}^{\Lambda}q_I^{-2}q_I^{d-1}dq_I,
	\qqq
	where $M$ is the contribution to the $q^2$ term of the equations of motion as reported above, and the subindex on the diagram refers to its second order contribution.

\section{Computation of the diagram in Fig. \ref{fig:diagrams}(f)}\label{app:1f}

	This diagram must be written, once we write specific momenta into the outcoming fields, as a sum of two diagrams in order to symmetrize it, so we write:
	\qq	
	\vcenter{\hbox{\includegraphics[page=2]{app_b}}}
	=
	\frac{1}{2}\left(\vcenter{\hbox{\vspace{0.17cm}\includegraphics[page=3]{app_b}}}+\;\;\vcenter{\hbox{\vspace{0.17cm}\includegraphics[page=4]{app_b}}}\right).
	\qqq
	
	In both diagrams we consider the correlator to carry the loop frequency $q_I$ and the propagator to carry the corresponding frequency to ensure momentum conservation. We have
	\qq
	\vcenter{\hbox{\includegraphics[page=1]{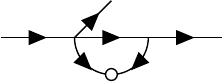}}}=&
	12\frac{1}{2}G_0(\hat{\bfq})\times\nonumber\\
	&\int_{q,q',q_I}\left[V_1G_0(\hat \bfq -\hat \bfq'-\hat \bfq_I)+V_2G_0(\hat \bfq'-\hat \bfq_I)\right]\times\nonumber\\
	&\qquad\quad C_0(\hat \bfq_I)\phi(\hat \bfq')\phi(\hat \bfq-\hat \bfq'),
	\qqq
	where the first factor of $12$ is the symmetry factor, the $1/2$ comes from the symmetrization above, the two fields at the end are the two external legs, and where $V_1$ and $V_2$ are the contributions from both diagrams in the symmetrization:
	\qq
	V_1&=\frac{-u}{2}\Bigg(& 2\lambda(-\bfq_I)\cdot(\bfq-\bfq') - \zeta\frac{q_I^2(\bfq-\bfq')\cdot(\bfq-\bfq'-\bfq_I)}{(\bfq-\bfq'-\bfq_I)^2}\nonumber \\
	&&-\zeta\frac{(\bfq-\bfq')^2(-\bfq_I)\cdot(\bfq-\bfq'-\bfq_I)}{(\bfq-\bfq'-\bfq_I)^2}+\nu(\bfq-\bfq'-\bfq_I)^2\Bigg)\\
	V_2&=\frac{-u}{2}\Bigg(&-2\lambda \bfq'\cdot \bfq_I -\zeta\frac{\bfq_I^2 \bfq'\cdot(\bfq'-\bfq_I)}{(\bfq'-\bfq_I)^2}\nonumber\\ &&-\zeta\frac{\bfq'^2(-\bfq_I)\cdot(\bfq'-\bfq_I)}{(\bfq'-\bfq_I)^2}+\nu (\bfq'-\bfq_I)^2\Bigg).
	\qqq
	Again the same strategy is followed as in \ref{app:1a}: we will expand in low external frequencies $q$ and $q'$ and collect terms. In this case, because the zero order term would produce a cubic term in the e.o.m. of the form of $\dot\phi=c\nabla^2\phi^2+\ldots$, that we are absorbing in the field via an additive renormalization, we study only the second term, that will contribute to the couplings. 
	
	We rewrite the scalar products by explicitly writing the angles between each pair of momenta. Unlike before, we know have three momenta so we need three angles, that we define as
	\qq
	\bfq\cdot \bfq'& \equiv q q' \cos\psi\\
	\bfq\cdot \bfq_I&\equiv q q_I \cos\theta \\
	\bfq'\cdot \bfq_I&\equiv q' q_I \cos\phi,.
	\qqq
	One last transformation has to be done to the angles. We will perform the internal frequency integral in spherical coordinates so we must transform these angles to those of this coordinate system. We consider $q$ to be in the $x_d$ axis, and $q'$ to be on the plane of the axes $x_d$ and $x_{d-1}$ (this can be done without losing generality by a rotation of the reference frame), and consider $q_I$ to be an arbtirary vector with the following spherical coordinates,
	\qq
	q_{I,1} &= q_I\sin\theta\ldots\sin\phi_{d-1}\sin\phi_{d-2}\\
	q_{I,2} &= q_I\sin\theta\ldots\sin\phi_{d-1}\cos\phi_{d-2}\\
	\vdots &\\
	q_{I,d-1} &=q_I\sin\theta\cos\phi_1\\
	q_{I,d} &=q_I\cos\theta.
	\qqq
	
	Here all angles vary in the interval $[0,\pi]$ except $\phi_{d-2}$ which varies in the interval $[0,2\pi]$. We observe that $\theta$ as defined is already appropriate for integration. We must therefore express $\phi$ in terms of the $\{\theta,\phi_1,...,\phi_{d-2}\}$, for which we use the following expression:
	\[
	q'q_I\cos\phi=\bfq'\cdot\bfq_I = q'q_I(\cos\theta\cos\psi+\sin\theta\cos\phi_1\sin\psi).
	\]
	
	The last step is performing the spherical integrals in the same way as we did in the \ref{app:1a}. This can be done term by term identifying trigonometric expressions with unitary vectors. The terms in this diagram will have the same integrals shown above, plus a new one, that can be calculated exactly as $\cos^2\theta$:
	\[
	\int_{\textrm{sphere}}\cos^2\phi_1 \sin^2\theta=\int u_{d-1}=\frac{1}{d}\frac{S_d}{(2\pi)^d}.
	\]
We then obtain	
	\qq
	\vcenter{\hbox{\includegraphics[page=1]{app_b}}}\,{}_{2}=&
	\frac{3Du\left[2(d-1)\zeta+(2-d)\nu\right]}{K^2d}\frac{S_d}{(2\pi)^d}\times\nonumber\\
	&\left(q^2+2q'^2-2q'q\cos\psi\right) \int_{\Lambda/b}^{\Lambda}q_I^{-4}q_I^{d-1}dq_I.
	\qqq
	
	Notice we have three terms, one proportional to $q^2$, one proportional to $q'^2$ and the last proportional to $\bfq\cdot \bfq'$. We have to absorb these three terms into the couplings $\nu$, $\zeta$ and $\lambda$. This diagram in particular will not contribute to $\zeta$ since with those terms we cannot build the neccessary dot products, so we can calculate the contributions by rewriting the general expression
	\[
	Aq^2+Bq'^2+C\bfq\cdot \bfq'=Aq^2+C\bfq'\cdot(\bfq-\bfq')+(B+C)q'^2,
	\]
	where the first term is absorbed by $\nu$, the second one by $\lambda$, and the third should be zero (and it trivially is) because the diagram must respects the symmetry of interchanging the two external legs which we imposed. Taking into account the prefactors in the equation of motion, $A$ and $C$ here are the terms $B_{2,\bar\nu}$ and $B_{2,\bar\lambda}$ in the flow equations.

\section*{References}
\bibliographystyle{unsrt}
\bibliography{biblio}

\end{document}